Evaluating the Impact of Bitcoin on International Asset Allocation using Mean-Variance, Conditional Value-at-Risk (CVaR), and Markov Regime Switching Approaches

Mohammadreza Mahmoudi

Department of Economics, Northern Illinois University, Dekalb, USA.

Email: mmahmoudi@niu.edu

Abstract

This paper aims to analyze the effect of Bitcoin on portfolio optimization using mean-variance, conditional value-at-risk (CVaR), and Markov regime switching approaches. I assessed each approach and developed the next based on the prior approach's weaknesses until I ended with a high level of confidence in the final approach. Though the results of mean-variance and CVaR frameworks indicate that Bitcoin improves the diversification of a well-diversified international portfolio, they assume that assets' returns are developed linearly and normally distributed. However, the Bitcoin return does not have both of these characteristics. Due to this, I developed a Markov regime switching approach to analyze the effect of Bitcoin on an international portfolio performance. The results show that there are two regimes based on the assets' returns: 1) bear state, where returns have low means and high volatility, 2) bull state, where returns have high means and low volatility.

# 1. Introduction

Friedrich Hayek, in his famous book "The Denationalization of Money", proposed that distrust in central banks leads to the abolishment of governmental monopoly in the supply of money and motivates the unregulated money market, which provides effective products for money-users and ameliorates the issues caused by business cycles Howard (1977). The main purpose of Blockchain technologies is record keeping in a decentralized manner and Bitcoin is the first and most famous realization of this idea.

There is disagreement among people regarding cryptocurrencies. Some people propose that it is innovation that helps ensure resources are allocated efficiently, while others maintain that it is an outright bubble, which only finances illegal activities around the world. Besides this profound disagreement, data indicates that cryptocurrencies have developed significantly since the creation of Bitcoin in 2009. As of March 2, 2022, there are over 10,000 cryptocurrencies with more than $2 trillion total market capitalization.[1] Acceptance, regulations, and participation of institutional investors increase the reliability of Bitcoin and the cryptocurrency industry as well.

Bitcoin, the world's first and most-popular cryptocurrency, solely accounts for nearly 41 percent of cryptocurrency total market capitalization. Hence, it has a deterministic effect on other cryptocurrencies' prices. As you see in Figure 1, the price of Bitcoin has fluctuated considerably. Its price increases from nothing to $67,559 in November 2021. It should be noted that the Bitcoin price jumped significantly every four year; The reasons behind this cyclical pattern of Bitcoin price is rooted in changes in supply and demand. The Bitcoin mining cost increases dramatically every four years, since the supply of Bitcoin is limited, the price goes up. This jump in price attracts

---

[1] The cryptocurrency market cap data available at https://coinmarketcap.com/charts/

more investors, motivates herding investment, and increases demand. Decreasing the supply and increasing the demand leads to a jump in Bitcoin price, which is known as cryptocurrency season. The latest cryptocurrency season, which started in late 2020 and continues to the present, is longer than the previous cryptocurrency seasons that occurred in 2013 and 2017. The latest season was generated by COVID-19 pandemic. COVID-19 had significant effects on the different parts of the US economy especially labor market and stock market Mahmoudi (2022). Following this evolution in the season length, many portfolio managers are now offering cryptocurrencies for risk-lover investors.

*Figure 1.Bitcoin (BTC) price*

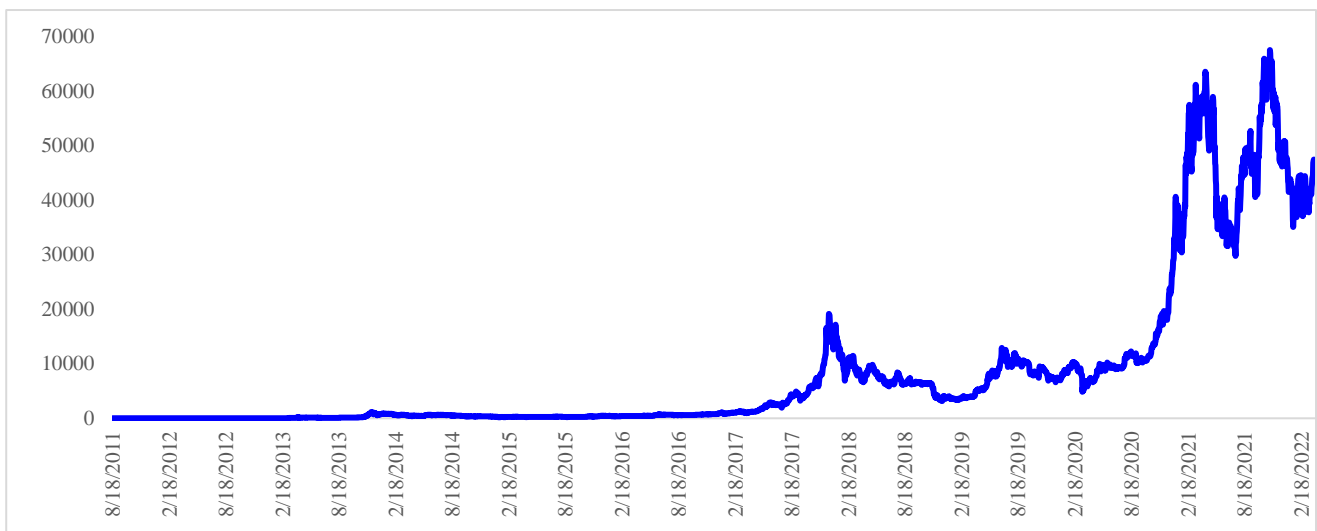

Note: This figure plots the historical data of Bitcoin (BTC) price from 08/18/2011 to 04/05/2022. As it is clear, every four years (2013, 2017, 2021) the Bitcoin price jumped dramatically. During these periods that famous as a cryptocurrency season due to the jump in Bitcoin mining cost the supply reduces significantly, then the demand increase because of herding and price increase dramatically.

This study aims to analyze the effect of adding Bitcoin to a well-diversified international portfolio using three main approaches: 1) Mean-variance, 2) Conditional Value-at-Risk (CVaR), 3) Markov regime switching. This paper will proceed as follows: Section 2 summarizes the main studies regarding portfolio optimization and the effect of cryptocurrencies on portfolio optimization;

Section 3 presents the relevant data sources; Section 4 explains the methodology of three different portfolio optimization analyses; and Section 5 concludes.

## 2. Literature Review

Bitcoin was invented in 2008 by an anonymous figure known as Nakamoto (2008), since then, there are plenty of studies that have analyzed the different aspects of Bitcoin. In the following I summarize the main papers that studied financial and economic aspects of Bitcoin.

Yermack (2013) studied whether Bitcoin behaves like a traditional currency using its historical trading behavior. The results show that Bitcoin behaves more like a speculative investment rather than like a currency. Cheung, Roca, and Su (2015) used an econometric approach named Phillips, Shi and Yu to detect bubbles in the price of Bitcoin from 2011-2014. They found three long-lasting bubbles, which endured from 66 days to 106 days. Bianchi (2020) used a data set of prices, traded volumes, and market capitalization for several cryptocurrencies to examine their relationship with traditional assets and derive the main factors behind their supply and demand. The results show that there is a significant relationship between cryptocurrencies' returns and commodities such as gold and energy.

Chiu and Koeppl (2017) examined the optimal design of cryptocurrencies by quantitatively evaluating how well such currencies can support bilateral trade. They estimated that the current Bitcoin platform produces a welfare loss of 1.4 percent of consumption, which could reduce to 0.08 percent by adopting an optimal design. Alexander and Dakos (2020) highlighted the importance of data resources of cryptocurrencies in order to get accurate results. They pointed out that more than half the papers published since January 2017 are using inaccurate cryptocurrency data. Tang and You (2021) found the sources of demand for Bitcoin using violations of the law of one price of Bitcoin. They maintained that there is a difference between the price of Bitcoin in local currency and average worldwide dollar Bitcoin prices, and this price deviation rises by growing in perceptions of institutional failures, increasing crypto-trading frictions, and creating cryptocurrency prices rallies. Also, the price deviation is stronger in the countries in which people express more distrust in each other.

There are some books and papers that gave a comprehensive perspective regarding technical and legal angles of cryptocurrencies like Franco (2014); Ali et al. (2014); Böhme et al. (2015); Narayanan et al. (2016); Halaburda and Sarvary (2016); Waelbroeck (2018); Raskin and Yermack (2018); Perkins (2018); Corbet et al. (2019); Treiblmaier and Beck (2019); Grabowski (2019), Mahmoudi (2021).

There are some studies that focus on the effect of Bitcoin on portfolio optimization. Using weekly data from 2011 to 2013, Briere, Oosterlinck, and Szafarz (2015) analyzed the effect of adding Bitcoin to a portfolio that includes both traditional assets like worldwide stocks, bonds, hard currencies and alternative investments like commodities, hedge funds, and real estate. They found that Bitcoin has high return and volatility, however, its correlation with other assets is low. They discovered that adding Bitcoin improves the risk-return trade-off of well-diversified portfolios. Eisl, Gasser, and Weinmayer (2015) considered a well-diversified portfolio, then asked how adding Bitcoin changed the risk and return of the portfolio by using mean conditional value-at-risk (mean-CVaR) portfolio optimization approach. The results indicated that Bitcoin increases the CVaR of a well-diversified portfolio, however, this additional risk is offset by Bitcoin's high return. Carpenter (2016) used a modified version of mean-variance portfolio optimization to show that Bitcoin can be a viable diversification tool, however, the results were affected by a speculative bubble that occurred in 2013.

Symitsi and Chalvatzis (2019) evaluated the out-of-sample performance of Bitcoin within a well-diversified portfolio under four strategies. They found significant diversification benefits of adding Bitcoin to a portfolio due to low correlation between Bitcoin and other assets. Also, they concluded that the non-bubble period leads to a dramatic reduction of the diversification benefits. Hu, Rachev, and Fabozzi (2019) maintained that the portfolios that were constructed by minimizing the variance and Conditional Value-at-Risk (CVaR) outperform the S&P 500 index. Also, they used a dynamic pricing model to price crypto asset options. Bakry et al. (2021) analyzed the performance of Bitcoin on portfolios using high sharp ratio criteria. Their results show that Bitcoin has safe haven properties and could act as a diversifier in normal market conditions. Moreover, using Markov regime switching approach Mahmoudi and Ghaneei (2022) detected two regimes in the Toronto Stock Exchange Index (TSX) based on monthly data from 1970 to 2021. Regime 1, where growth rate of stock index is positive; and regime 2, where growth rate of stock index is negative.

## 3. Data

In this study I use monthly data for the Bitcoin price (BTC) from July, 2010 to April, 2022.[2] Moreover, the international asset allocation analysis incorporates the data of the world market portfolio (MSCI World), and the largest international stock markets, namely the United States (MSCI United States), the United Kingdom (MSCI United Kingdom), continental Europe (MSCI Europe ex-UK), Japan (MSCI Japan), and the Pacific region (MSCI Pacific ex-Japan) for the same period of time.[3] Furthermore, the risk-free rate is measured by the 30-day US Treasury bill rate provided by the federal reserve.[4]

---

[2] The Bitcoin price data available at https://www.tradingview.com/chart/xHhn8Bhu/

For portfolio optimization analysis, I used real asset returns, which are computed as follows:

- I computed nominal asset return by using the formula $\frac{P_t - P_{t-1}}{P_{t-1}}$, where $P_t$ is the price of assets at time t.

- Then, I used inflation rate to convert the nominal return to real return using the formula $\frac{1+r_{it}}{1+i_t}$, where $r_{it}$ is the nominal asset return at time t and $i_t$ is the inflation rate at time t, which was computed based on consumer price index (CPI) for all urban consumers.

Some of the main statistics of Bitcoin and international stock real returns are summarized in Table 1. Bitcoin's return fluctuated significantly, and it has higher mean and standard deviation compared to other assets. These results confirm the risk-return tradeoff principle[5] as shown in Figure 2 as well.

Table 1. Descriptive statistics of Asset returns

|  | Mean | Median | Standard Deviation | Kurtosis | Skewness | Range | Minimum | Maximum | Count |
|---|---|---|---|---|---|---|---|---|---|
| BTC | 1.172038733 | 1.055794891 | 0.560214599 | 17.90606008 | 3.771283261 | 4.156613063 | 0.613933829 | 4.770546892 | 141 |
| MSCI World | 0.482111888 | 0.490679174 | 0.079260998 | 3.10341471 | -1.123099522 | 0.506386593 | 0.17676836 | 0.683154953 | 141 |
| MSCI USA | 0.683637294 | 0.679952167 | 0.044266323 | 2.810852507 | 1.063725771 | 0.287605011 | 0.583489054 | 0.871094066 | 141 |
| MSCI UK | 0.595941886 | 0.597575714 | 0.022178124 | 0.520149899 | -0.261825374 | 0.120986453 | 0.536048537 | 0.65703499 | 141 |
| MSCI Europe Ex UK | 0.629781454 | 0.63214637 | 0.021626692 | 0.757571116 | -0.573369791 | 0.134229522 | 0.558352458 | 0.69258198 | 141 |
| MSCI Japan | 0.617580191 | 0.617728963 | 0.023967518 | 0.135396955 | -0.039075547 | 0.119771629 | 0.561928095 | 0.681699724 | 141 |
| MSCI Pacific | 0.622059806 | 0.623335847 | 0.029836927 | 2.367705242 | -0.375413132 | 0.213017694 | 0.501491076 | 0.71450877 | 141 |

---

[3] The international stock index data available at https://www.msci.com/real-time-index-data-search
[4] The 30-day US Treasury bill rate available at https://fred.stlouisfed.org/series/DGS1MO
[5] The risk-return tradeoff expresses that the greater the risk on an optimal portfolio, the greater the expected return on that portfolio LeRoy and Werner (2014)

| | | | | | | | | |
|---|---|---|---|---|---|---|---|---|
| **Ex Japan** | | | | | | | | |
| **US 1-Month T-bills** | 0.795015193 | 0.616290363 | 1.657353951 | 30.57653758 | 3.186007669 | 20.98877475 | -7.693696899 | 13.29507785 | 141 |

Note: This table summarized some of the main descriptive statistics of returns of Bitcoin, MSCI world, MSCI USA, MSCI UK, MSCI Europe ex UK, MSCI Japan, MSCI Pacific ex Japan, and US 1-month treasury bills. The results indicates that Bitcoin real return has highest average and standard deviation compared to other assets.

*Figure 2.Real Returns of Bitcoin and International assets*

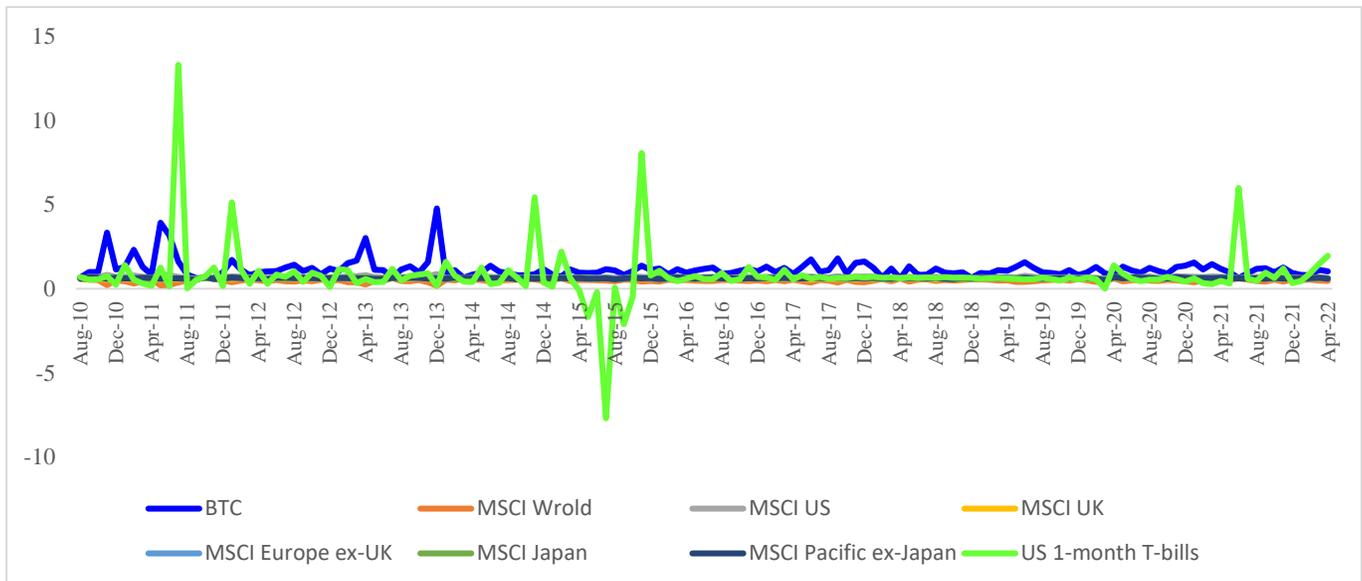

Note: This figure depicts the historical real returns of different asset. Bitcoin return has the highest fluctuations. I computed nominal return by using the formula $\frac{P_t - P_{t-1}}{P_{t-1}}$, where $P_t$ is price of assets at time t. Then, I use inflation rate to convert the nominal return to real return using the formula $\frac{1+r_{it}}{1+i_t}$, where $r_{it}$ is the nominal asset return i at time t and $i_t$ is inflation rate at time t, which computed based on consumer price index (CPI) for all urban consumers. The graph show that Bitcoin return is highly fluctuated compared to other stocks.

The MSCI World Index with 1,540 constituents is designed to capture large and mid-cap representation across 23 Developed Markets countries. This index represents approximately 85% of the free float-adjusted market capitalization in each country. Moreover, the MSCI USA Index with 627 constituents, the MSCI United Kingdom Index with 81 constituents, the MSCI Japan Index with 260 constituents, the MSCI Europe ex-UK Index with 348 constituents, and the MSCI Pacific ex-Japan Index with 121 constituents are designed to measure the performance of the large and mid-cap segments and cover 85% of the free float-adjusted market capitalization in the US,

UK, Japan, 14 Developed Markets countries in Europe (excluding UK), and 4 of 5 Developed Markets countries in the Pacific region (excluding Japan).[6]

The correlation coefficients between assets are shown in **Error! Reference source not found.**. Based on the results, MSCI World and Bitcoin returns are highly negatively correlated. Also, the correlation coefficient between MSCI UK and Bitcoin is around -0.59, which is high. However, MSCI US is highly positively correlated with Bitcoin. The correlation of MSCI Europe ex-UK and MSCI Japan with Bitcoin is 0.21 and 0.18 respectively. The correlation coefficients of MSCI Pacific ex-Japan and US 1-month treasury bill with Bitcoin are small and around -0.05 and 0.04, respectively.

Table 2: Correlation coefficients of assets' returns

| | BTC | MSCI World | MSCI US | MSCI UK | MSCI Europe ex-UK | MSCI Japan | MSCI Pacific ex-Japan | US 1-month T-bills |
|---|---|---|---|---|---|---|---|---|
| BTC | 1 | | | | | | | |
| MSCI World | -0.899410844 | 1 | | | | | | |
| MSCI US | 0.862045754 | -0.787293815 | 1 | | | | | |
| MSCI UK | -0.585385261 | 0.716525815 | -0.422379548 | 1 | | | | |
| MSCI Europe ex-UK | 0.215228646 | -0.049546817 | 0.540784214 | 0.219775739 | 1 | | | |
| MSCI Japan | 0.183992924 | -0.083852555 | 0.38217477 | 0.147872605 | 0.527447299 | 1 | | |
| MSCI Pacific ex-Japan | -0.051965578 | 0.203249143 | 0.253690658 | 0.424709816 | 0.524463447 | 0.124163037 | 1 | |
| US 1-month T-bills | 0.04138583 | -0.055290322 | 0.05110286 | -0.037710118 | 0.022530494 | 0.040380562 | 0.016517151 | 1 |

Note: This table shows the correlation coefficient between assets' returns. The results show that Bitcoin and MSCI world are highly negatively correlated. While Bitcoin and MSCI world are highly positively correlated. Moreover, MSCI UK and Bitcoin are negatively correlated. MSCI Europe ex-UK and MSCI Japan are positively correlated with Bitcoin. Furthermore, the correlation coefficient MSCI Pacific ex-Japan and US 1-month treasury bill with Bitcoin are small and around -0.05 and 0.04, respectively.

---

[6] The information about different MSCI index is available at https://www.msci.com/

## 4. Methodology

I wanted to analyze the effect of adding Bitcoin to a well-diversified international portfolio. To accomplish this, I used three different portfolio optimization analyses: 1) Mean-variance, 2) Conditional Value-at-Risk (CVAR), 3) Markov regime switching. In the following subsections, I evaluated each analysis and developed the next based on the weaknesses of the prior analysis until I ended with a high level of confidence in the final portfolio optimization analysis.

### 4.1 Mean-Variance Approach

Mean-variance analysis was introduced by Markowitz (1952). It is a mathematical framework of portfolio optimization such that the risk is minimized for a given level of expected return. I used the Cochrane (2009) notations to develop the mean-variance approach as follows:

$$min_{\{w\}} w'\Sigma w$$

$$\text{s.t } w'E = \mu;$$

$$w'1 = 1$$

where w is assets' weights, E denotes vector of mean returns ($E \equiv E(R)$), and Σ denotes variance-covariance matrix ($\Sigma = E\text{Error! }\boldsymbol{Bookmar}$k not defined.).

The solution is as follows:

Let

$$A = E'\Sigma^{-1}E; \quad B = E'\Sigma^{-1}1; \quad C = 1\Sigma^{-1}1$$

for a given portfolio mean return ($\mu$), the minimum variance portfolio has variance:

$$var(R^p) = \frac{C\mu^2 - 2B\mu + A}{AC - B^2}$$

Then, the optimal weights are:

$$w = \frac{\Sigma^{-1}1}{1'\Sigma^{-1}1}$$

The optimal weights of mean-variance portfolio for our data are shown in Figure 3. Based on the results, Bitcoin has the highest weight (31.5%) in mean-variance portfolio. After Bitcoin, MSCI Japan (23%), US 1-month treasury bills (19.4%), and MSCI US (17.9%) have the second, third, and fourth highest weights. Optimal weight of MSCI Europe ex-UK and MSCI Pacific ex-Japan are 6.1 and 2.1 precent, respectively. Moreover, MSCI World and MSCI UK have zero weights in our optimal portfolio. Based on these optimal weights, the return and standard deviation of optimal portfolio are 83 percent and 37 percent, respectively.

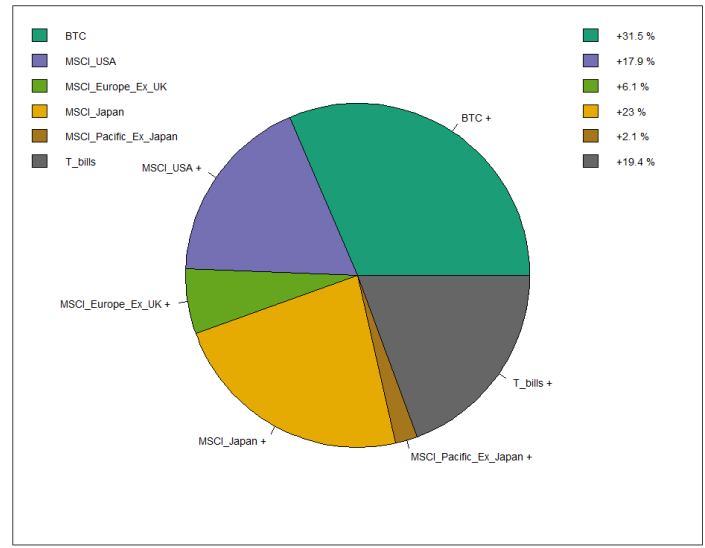

*Figure 3.Optimal Weights of Mean-variance Portfolio*

Note: This figure shows the optimal weights based on mean-variance portfolio optimization approach. Based on the results Bitcoin has highest weight (31.5 percent) in mean-variance portfolio. After Bitcoin, MSCI Japan, US 1-month treasury bills, and MSCI US have the highest weights. Moreover, optimal weight of MSCI world, MSCI UK, MSCI Europe ex UK, MSCI Pacific ex Japan are 0, 0, 6.1, 2.1 precents respectively.

The main drawback of the Mean-variance portfolio optimization approach is that the asset returns should be normally distributed McNeil, Frey, and Embrechts (2015), however, the Bitcoin return does not follow normal distribution.

## 4.2   Conditional Value-at-Risk (CVaR) Approach

Bitcoin price fluctuated considerably. We need an approach for portfolio optimization to reduce the risk of high loss. Value at risk (VaR) calculations depend on linear estimation of the portfolio risks and assume a joint normal or log-normal distribution of returns. Also, it has undesirable mathematical features like lack of convexity and subadditivity. An alternative to VaR is Conditional value-at-risk (CVaR), which also measures the risk and has some dominant characteristics like transition-equivariance, convexity, and positively homogeneity. Portfolios with low CVaR automatically lead to low VaR as well Rockafellar and Uryasev (2000). Since

Bitcoin return is not normally distributed, I adopted the following CVaR framework to analyze the impact of Bitcoin on the performance of a well-diversified international portfolio

For the CVaR approach, I use the notations introduced by Krokhmal, Palmquist, and Uryasev (2002). Let $f(w,y)$ be the loss related to the decision vector w (i.e., portfolio weights) and the random vector y (i.e., uncertainties like market variables that affect the loss). We could write the cumulative distribution function for the loss associated with w as follows:

$$\Psi(w,\varsigma) = P\{y|f(w,y) \leq \varsigma\} = \int_{f(w,y)\leq\varsigma} p(y)\,dy$$

Where $\varsigma$ is a specific level of risk and $\Psi(w,\varsigma)$ is nondecreasing and a continuous function with respect to $\varsigma$.

The value at risk (VaR) for specific probability level α is:

$$\varsigma_\alpha(w) = \min\{\varsigma|\Psi(w,\varsigma) \geq \alpha\}$$

Moreover, the conditional value at risk (CVaR) for the loss random variable associated with w and specific probability level α is:

$$CVaR_\alpha(w) = \frac{1}{1-\alpha} \int_{f(w,y)\geq\varsigma_\alpha(w)} f(w,y)p(y)\,dy$$

Then, we could write the portfolio optimization problem as follows:

$$min_{\{w\}} CVaR_\alpha(w)$$

$$\text{s.t } w'E = \mu;$$

$$w'1 = 1$$

I evaluated the effect of adding Bitcoin to a well-diversified international portfolio under CVaR approach using the following portfolio strategies:

1) Unconstrained short portfolio: This strategy does not consider any weight-related constraints in a portfolio optimization problem.

2) Long only portfolio: This strategy does not consider short sell. Hence, the weights are bounded between 0 and 1.

3) Box-constrained portfolio: This strategy considers upper and lower bounds on the asset weights.

4) Equally-weighted portfolio: The assets' weights are equal and constant over time.

The average Bitcoin weight, portfolio return, CVaR, and risk-return ratio for these four strategies are shown in Table 2. Based on the results, the average Bitcoin weight is small with minimum quantity of 1.65 percent in box-constrained strategy and maximum quantity of 7.8 in equally-weighted strategy. It should be noted that in all strategies, the average return of the portfolio with Bitcoin is significantly greater than the average return of the portfolio without Bitcoin. The maximum average return belongs to the unconstrained strategy of a portfolio with Bitcoin. Also, the average CVaR criteria of a portfolio with and without Bitcoin for different strategies indicates that portfolio risk rises when Bitcoin is added to the portfolio. The maximum CVaR quantity belongs to the unconstrained strategy of a portfolio with Bitcoin. Most notably, risk-return ratio confirms that higher average return of a portfolio with Bitcoin overcompensates the rise in risk.

*Table 2. Portfolio optimization strategies under Conditional Value-at-Risk (CVaR)*

|  | Unconstrained short $w_i \in R$ | | Long only $w_i \geq 0$ | | Box constrained $-1 \leq w_i \leq 1$ | | Equally weighted $w_i = 1/N$ | |
| --- | --- | --- | --- | --- | --- | --- | --- | --- |
|  | Without Bitcoin | With Bitcoin | Without Bitcoin | With Bitcoin | Without Bitcoin | With Bitcoin | Without Bitcoin | With Bitcoin |
| **Average Bitcoin Weight** | - | 6.7 | - | 2.1 | - | 1.65 | - | 7.8 |
| **Average Return** | 0.44 | 5.45 | 0.24 | 0.56 | 0.44 | 1.05 | 0.42 | 1.98 |

| Average CVaR | 0.34 | 1.98 | 0.5 | 0.65 | 0.33 | 0.57 | 0.6 | 1.21 |
| --- | --- | --- | --- | --- | --- | --- | --- | --- |
| **Average risk-return Ratio** | 1.8 | 5.83 | 0.6 | 1.14 | 1.84 | 2.88 | 2.2 | 3.8 |

Note: This tables summarizes the main results of four strategies of portfolio optimization under Conditional Value-at-Risk (CVaR). The results show that the average Bitcoin weight is small with minimum quantity of 1.65 percent in Box constrained strategy and maximum quantity of 7.8 in equally weighted strategy. In all strategies the average return of portfolio with Bitcoin is considerably more than the average return of portfolio without Bitcoin. The maximum average return belongs to unconstrained strategy of a portfolio with Bitcoin. Also, average CVaR criteria of portfolio with and without Bitcoin for different strategies indicate that portfolio risk goes up when Bitcoin add to the portfolio. The maximum CVaR quantity belongs to unconstrained strategy of a portfolio with Bitcoin. However, the risk-return ratio confirms that higher average return of portfolio with Bitcoin overcompensate the rise in risk.

The main conclusion that we can deduct from the above approaches is that Bitcoin helps the diversification of an international portfolio. However, their main drawback is that they rely on the historical data. Therefore, the computed assets' weights do not have predictive power. In response to this situation, newer approaches like portfolio optimization under Markov regime switching can capture the nonlinear relations between assets' returns and provide more predictive and applicable assets allocation. That being said, I want to evaluate the asset allocation under different regimes to give investors and asset managers a predictive power that is based on regimes' probabilities.

## 4.3 Markov Regime Switching Approach

The majority of portfolio optimization literature assumes that a linear process generates asset returns, while empirical evidence indicates that asset returns have a more complicated dynamic with multiple regimes Guidolin and Timmermann (2007). Regime switching models can detect regimes with very distinctive mean, variance, and correlations across assets. Also, they are able to capture complicated forms of heteroskedasticity, fat tails and skews in the returns distribution Timmermann (2000).

To be specific, consider the standard power utility function

$$u(c) = \frac{c^{1-\gamma}}{1-\gamma}$$

Let $g_{t+1} = \frac{c_{t+1}}{c_t}$ be a consumption growth. Then, we could calculate the stochastic discount factor as follows:

$$m_{t+1} = \beta \frac{U'(c_{t+1})}{U'(c_t)} = \beta \frac{c_{t+1}^{-\gamma}}{c_t^{-\gamma}} = \beta \left(\frac{c_{t+1}}{c_t}\right)^{-\gamma} = \beta (g_{t+1})^{-\gamma}$$

The expected (gross) return between periods t and t + 1, $r_{i,t+1}$ for any asset i; satisfies

$$1 = E_t m_{t+1} r_{i,t+1}$$

The risk premium is then given by

$$E_t \text{Error! Bookmark not defined.} = -\frac{Cov_t[m_{t+1}, (r_{i,t+1} - r_t^f)]}{E_t[m_{t+1}]}$$

Suppose consumption growth rate follows a regime switching process in which both mean and variance of $g_{t+1}$ depends on the state of the economy (i.e. $g_{t+1} \sim N(\mu_{s_{t+1}}, \sigma^2_{s_{t+1}})$; where $s_{t+1} = 1,\ldots,k$ denotes the state of the economy ). This indicates that the stochastic discount factor follows a k-state process. Therefore, we could write the risk premium as follows:

$$E_t[r_{i,t+1} - r_t^f] = -\frac{\sum_{s_{t+1|t}}^{k} \pi_{s_{t+1|t}} Cov[m_{t+1}, (r_{i,t+1} - r_t^f)|s_{t+1}]}{\sum_{s_{t+1|t}}^{k} \pi_{s_{t+1|t}} E[m_{t+1}|s_{t+1}]}$$

Where $\pi_{s_{t+1|t}}$, is state probability at time t+1 given information at time t. Thus, risk premium follows a regime switching process driven by states as well.

In order to compute the regimes' probabilities of the joint distribution of asset returns and predictor variables, I consider $(n + m) \times 1$ vector of excess assets' returns, $r_t = (r_{1t}, r_{2t}, \ldots, r_{nt})'$ extended by a set of m predictor variables $z_t = (z_{1t}, z_{2t}, \ldots, z_{mt})'$. Suppose that the mean, covariance and serial correlations of asset's returns follow a common state variable, $S_t = 1, \ldots, k$ Then the dynamics of $r_t$ and $z_t$ are given by:

$$\begin{bmatrix} r_t \\ z_t \end{bmatrix} = \begin{bmatrix} \mu_{S_t} \\ \mu_{zS_t} \end{bmatrix} + \sum_{j=1}^{P} A_{j,S_t} \begin{bmatrix} r_{t-j} \\ z_{t-j} \end{bmatrix} + \begin{bmatrix} \varepsilon_{r_t} \\ \varepsilon_{z_t} \end{bmatrix}$$

Where, $\mu_{S_t}$ and $\mu_{zS_t}$ denote the intercept vectors of $r_t$ and $z_t$ in state $S_t$. $\{A_{j,S_t}\}_{j=1}^{P}$ denotes $(n+m) \times (n+m)$ matrices of autoregressive coefficients in state $S_t$. $\begin{bmatrix} \varepsilon_{r_t} \\ \varepsilon_{z_t} \end{bmatrix} \sim N(0, \Sigma_{st})$, where $\Sigma_{st}$ is a $(n+m) \times (n+m)$ covariance matrix.

In the above equations, we assumed the state variables, $s_t$, which specify the switching behavior of time series variables, follows an irreducible ergodic two-state Markov process. This assumption indicates that a current regime $s_t$ depends on the regime one period ago, $s_{t-1}$ Krolzig (1998). Hence, the transition probability between states is as follows:

$$Pr(S_t = j | S_{t-1} = i, S_{t-2} = k, \ldots) = Pr(S_t = j | S_{t-1} = i) = p_{ij}$$

$p_{ij}$ denotes the transition probability from state $i$ to state $j$.

Generally, the transition probability is identified by a $(n \times n)$ matrix as follows:

$$P = \begin{bmatrix} p_{11} & p_{12} & \cdots & p_{1m} \\ p_{21} & p_{22} & \cdots & p_{2m} \\ \vdots & \vdots & \ddots & \vdots \\ p_{m1} & p_{m2} & \cdots & p_{mm} \end{bmatrix}$$

$$\sum_{j=1}^{k} p_{ij} = 1, i = 1,2,\ldots, k, \text{and } 0 \leq P_{ij} \leq 1$$

This general model allows means, variances and correlations of assets' returns to vary across states. Therefore, risk-return trade-off is different across states in a way that may have strong asset allocation implications. For instance, a risky asset like Bitcoin is a more attractive investment opportunity in the bull states rather than bear states. The model helps investors with previous risk aversion parameter to change assets allocation by knowing that the current state is bear or bull.

## 5. Conclusion

I started the analysis of the impact of adding Bitcoin to a well-diversified international portfolio by using mean-variance framework. The results show that Bitcoin has the highest weight in the optimal portfolio (around 31%) and could improve the diversification of the portfolio. Based on mean-variance assumption, assets' returns should be normally distributed. However, since the Bitcoin return does not follow normal distribution, the results of mean-variance approach may be misleading. Then, I adopted a conditional value-at risk (CVaR) portfolio optimization analysis, which was compatible with non-normal distributions of assets' returns. Based on the results, Bitcoin weight is small (the average weight is 5%) in the optimal portfolio. Although adding Bitcoin to a well-diversified international portfolio increased the CVaR, the high return of Bitcoin leads to overcompensating the risk and better risk-return ratio. It should be noted that non-linearity of asset's returns, especially Bitcoin returns, cast doubt on the CVaR results. Due to this, I adopted a Markov switching approach that covers the optimization portfolio cases where a non-linear process generates asset returns. The results indicate two regimes in asset's returns: 1) bear regime where assets' returns have low means and high volatility, and 2) bull regime where assets' returns have high means and low volatility. The weight of Bitcoin is different in each regime.

# Refrences